*Article*

# A Ti/Pt/Co multilayer stack for transfer function based magnetic force microscopy calibrations

**Baha Sakar [1,2,*], Sibylle Sievers [2], Alexander Fernández Scarioni [2], Felipe Garcia-Sanchez [3], İlker Öztoprak [1], Hans Werner Schumacher [2] and Osman Öztürk [1]**

[1] Gebze Technical University, Department of Physics, 41400, Kocaeli Turkey
[2] Physikalisch-Technische Bundesanstalt, 38116 Braunschweig, Germany
[3] Dpto. Física Aplicada, University of Salamanca, 37008 Salamanca, Spain
* Correspondence: bsakar@gtu.edu.tr; Tel.: (+90) 535 0434006

**Abstract:** Magnetic force microscopy (MFM) is a widespread technique for imaging magnetic structures with a resolution of some 10 nanometers. MFM can be calibrated to obtain quantitative (qMFM) spatially resolved magnetization data in units of A/m by determining the calibrated point spread function of the instrument, its instrument calibration function (ICF), from a measurement of a well-known reference sample. Beyond quantifying the MFM data, a deconvolution of the MFM image data with the ICF also corrects the smearing caused by the finite width of the MFM tip stray field distribution. However, the quality of the calibration depends critically on the calculability of the magnetization distribution of the reference sample. Here, we discuss a Ti/Pt/Co multilayer stack which shows a stripe domain pattern as a suitable reference material. A precise control of the fabrication process combined with a characterization of the sample micromagnetic parameters allows to reliably calculate the sample's magnetic stray field, proven by a very good agreement between micromagnetic simulations and qMFM measurements. A calibrated qMFM measurement using the Ti/Pt/Co stack as a reference sample is shown and validated and the application area for quantitative MFM measurements calibrated with the Ti/Pt/Co stack is discussed.

## 1. Introduction

MFM is a versatile tool for imaging magnetic nanostructures [1] which is available in many laboratories. Highest resolutions, down to 10 nm have been reported [2-4]. In MFM, a magnetically coated microscopic tip on a cantilever is scanned over a magnetic sample at a certain distance. In the dynamic mode of MFM, the cantilever oscillates and the interaction of the magnetic tip with the magnetic sample is monitored as a phase shift of the cantilever oscillation. This yields an initially qualitative image of the domain pattern of the magnetic sample. However, after a calibration, MFM can also provide quantitative images of magnetic fields or magnetization patterns in units of A/m. Calibration concepts discussed in literature can be separated in two categories: point-probe models and transfer function approaches. Point-probe models describe in simplified form of the magnetic tip as a point like [5,6] or extended [7] magnetic dipole/and or monopole, whose position and strength is found from a measurement of a reference sample. As reference samples for point-probe calibrations, micrometer-scale current rings [8] and parallel wires [9] or hard disks [10-13], and recently superconducting flux quanta [14] have been proposed. However, point-probe models show a strong dependence on the feature size. This can be overcome by fully considering the non-local structure of the tip, which can be done through use of a transfer function (TF) approach [15,16]. The TF approach regards the extended nature of the tip by determining the full point spread function of the instrument. In calibrated measurements, the TF approach corrects for the smearing caused by the extended tip and allows for high resolution quantitative magnetic field imaging, as can be seen in recent publications on, e.g. the determination of the local Dzyaloshinskii-Moriya interaction in skyrmion systems [17], characterizations of the stray field landscape of asymmetric Néel walls [18] or the observation of distinct skyrmion phases in hybrid ferromagnetic/ferrimagnetic multilayers [19]. TF calibrated MFM opens the opportunity to analyze nanoscale structures on millimeter sized imaging areas [20]. The calibration is, as are calibrations for point-probe models, based on a reference sample measurement. However, in this case, the reference measurement serves to determine the instruments transfer function in Fourier space, i.e. the wave vector dependent sensitivity of the instrument. Therefore, TF based calibrations require specific reference samples that have to meet stringent requirements with respect to the covered spatial structure spectrum, the invasiveness with respect to the tip magnetization distribution and, in particular, the calculability of the sample spatial magnetization or field distribution. As a consequence, reference samples for TF calibrations are scarcely available and only a few types of reference sample have been discussed in literature, so far: In 1998, a first set of reference samples based on different stacks of $Co_xNi_{1-x}$ and Pt layers was introduced to compare the resolution of different MFM setups [21]. Other reference samples are exploiting intrinsic domain patterns in different stacks (Cu(200 nm) Ni/Cu/Si(001) [16] and Co/Pt multilayer [22,23]) or rely on a pattern written on a hard disk [24].

To determine the specific requirements for MFM calibrations and to assess the suitability of a reference sample, the detailed imaging process and signal generation process needs to be understood. The data obtained in an MFM measurement is the phase shift $\Delta\Phi$ of the oscillation of the magnetically coated tip, which results from the interaction of the tip magnetic stray field with the magnetization distribution of the sample at a certain measurement height. $\Delta\Phi$ depends on the z-component of the tip's magnetic field $\mu_0 H_z^{tip}$ at the sample surface, the effective magnetic surface charge density $\sigma_{eff}$ (or, equivalently, the magnetization pattern) of the sample, and the mechanical properties of the cantilever, namely its quality factor $Q$ and the cantilever stiffness $c$. The relation between $\Delta\Phi$ and $\sigma_{eff}$ can conveniently be described in partial Fourier space. Here and in the following, bold letters denote vector quantities. To this end, the $x$ and $y$ coordinates are transferred into Fourier space $(x, y) \to (k_x, k_y) = \boldsymbol{k}$ and $k = \sqrt{k_x^2 + k_y^2}$ while the z-component is retained [16,25]:

$$\Delta\Phi(\boldsymbol{k}, z') = \sigma_{eff}^{*}(\boldsymbol{k}) \cdot \frac{Q}{c} \cdot [LCF(\boldsymbol{k}, \theta, A)]^2 \cdot k \cdot \mu_0 H_z^{tip}(\boldsymbol{k}, 0) = \sigma_{eff}^{*}(\boldsymbol{k}) \cdot ICF(\boldsymbol{k}, \theta, A) \quad (1)$$

The lever correction function $[LCF(\boldsymbol{k}, \theta, A)]^2$ corrects the impact of the polar angle of the cantilever with surface normal $\theta$ and the finite cantilever oscillation amplitude $A$, respectively. The asterisk indicates the complex conjugate. In partial fourier space, a multiplication with $-k$ gives the z-derivative, and thus $-k \cdot \mu_0 H_z^{tip} = \mu_0 \frac{dH_z^{tip}}{dz}$, which is the tip's stray field gradient, which is also called Tip transfer function, $TTF$, comprising all magnetic properties of the tip. The mechanical properties together with the $TTF$ give the so-called instrument calibration function ($ICF$), which is the calibrated point spread function of the specific MFM setup. The $ICF$ can be determined from a calibration measurement $\Delta\Phi^{ref}$ of a well-known reference effective charge distribution $\sigma_{eff}^{ref}$ by a regularized deconvolution using a pseudo-Wiener filter in the form:

$$ICF(\boldsymbol{k}, z') = \Delta\Phi^{ref}(\boldsymbol{k}, z') \cdot \frac{\sigma_{eff}^{ref*}(\boldsymbol{k})}{\left|\sigma_{eff}^{ref*}(\boldsymbol{k})\right|^2 + \alpha} \quad (2)$$

The regularization parameter $\alpha$ can be optimized by applying an L-curve criterion [26].

Vice versa, once the $ICF$ is known, and with the measurement parameters kept constant, a calibrated measurement of a sample under test (SUT) can be performed yielding the SUT's effective charge density distribution from the measured phase shift data $\Delta\Phi^{SUT}$ (with a second, L-curve optimized regularization parameter $\alpha'$):

$$\sigma_{eff}^{SUT} = \Delta\Phi^{SUT}(k, z') \cdot \frac{ICF^{*}(\boldsymbol{k}, z')}{|ICF(\boldsymbol{k}, z')|^{2} + \alpha'} \tag{3}$$

Note that while the cantilevers mechanical properties $Q$ and $c$ are required to calculate the tip's $TTF$, they need not be known or considered for calibrated charge density measurements if they keep unchanged, since they are embedded in the $ICF$ and thus are implicitly regarded in the calibration process. The cantilever stiffness $c$ is supposed to be constant, however, if of interest for the TTF calculation, can be characterized by e.g. fast in-situ "Thermal Noise" methods (rel. uncertainty 7-10% [27, 28]), or using MEMS (rel. uncertainty 7%), or microbalance (rel. uncertainty 4%) based stiffness measurement techniques [28-30]. However, the calibration step presented in Eq. 2 critically depends on how well the reference effective charge density distribution $\sigma_{eff}^{ref}$ is known. Due to a lack of other quantitative imaging techniques with comparable resolution that can be used to trace back magnetization distribution of a reference sample, a suitable reference sample needs to be calculable. This requires a deep understanding of its micromagnetic properties. A calculability is, in particular, given for the special case of thickness-independent strictly perpendicular magnetization structures with either up or down magnetized domains and well-defined domain transitions. Such a sample allows to calculate the effective magnetic surface charge $\sigma_{eff}^{ref}$ from the measured MFM phase shift image using the material parameters saturation magnetization $M_s^{ref}$, total magnetic layer thickness $d^{ref}$, and domain wall width $\delta_{DW}^{ref}$. To this end, up- and down-magnetized areas of the sample are identified by a discrimination based on a threshold criterion resulting in a binary matrix (1, -1). This is followed by a convolution with a domain wall operator [31,32] to introduce domain wall transitions, resulting in the normalized magnetization distribution $m(\boldsymbol{x})$ in real, or $m(\boldsymbol{k})$ in Fourier space. From the latter, $\sigma_{eff}^{ref}$ can be calculated by using the relation [16]:

$$\sigma_{eff}^{ref} = m(\boldsymbol{k}) \cdot M_s \cdot \left(1 - e^{-\boldsymbol{k}d}\right) \tag{4}$$

In this manuscript, we present a perpendicularly magnetized Ti/Pt/Co multilayer stack thin film material as a suitable reference sample candidate for TF based calibrations and will demonstrate that it fulfills the requirements discussed above. Well defined, high quality interfaces combined with a controlled fabrication process ensure a well-known stack geometry with low surface roughness. The stack shows a stripe domain pattern which gives a broad spatial feature spectrum. The Ti interlayer allows to reduce the averaged magnetization compared to Co/Pt stacks, so that a stray field amplitude <60 mT is found at a measurement height of 64 nm, typical for thin film characterizations, making it a suitable reference sample material also for low coercivity tips. Micromagnetic simulations reveal the magnetization structure and domain wall characteristics of the stack, which in combination with the highly reproducible fabrication process make the properties of the stack calculable. This is validated by a comparison with an existing reference sample. The application area in terms of accessible magnetic structure sizes covered by a calibration using the Ti/Pt/Co stack is discussed.

## 2. Results

### 2.1. Fabrication of the multilayer stack

The Ti/Pt/Co multilayer stack is grown by using magnetron sputtering with a layer architecture of [Ti(0.3 nm) / Pt(1.5 nm) / Co(0.5 nm) / Pt(1.5 nm)]$_{20}$ on a naturally oxidized Si(111) wafer. This sample will be referred to as Ti/Pt/Co or tpc in the remainder of this paper. The Ti, Pt and Co layers were deposited by using pulsed DC, DC and RF power sources respectively. The deposition chamber's base pressure was $3 \times 10^{-9}\ mbar$. The substrates were annealed prior to the deposition to clean it from residual surface contaminations as carbon and oxygen. The purity of the substrate and the targets were checked by x-ray photoemission spectroscopy (XPS). The XPS system, mounted to the same UHV cluster as the deposition system, allows to control the quality of the deposition. The deposition rates were calibrated by using XPS prior to the deposition and monitored by a quartz crystal microbalance (QCM) during the deposition. The QCM is calibrated according to the calibration values obtained from XPS. The calibrated deposition rates are $0.018\ nms^{-1}$ , $0.019\ nms^{-1}$ and $0.037\ nms^{-1}$ for Ti, Pt and Co respectively. As a result, the layer thicknesses are traceably defined and very reproducible. Alongside the thickness calibration, XPS based monitoring of the fabrication allows to reproduce the sample interface and layer structure in further depositions with very high accuracy. Furthermore, as an advantage of magnetron sputtering technique, samples can be prepared on substrates with radii up to 5 cm. This guarantees a high availability of the reference material. The detailed steps for the calibration of deposition and XPS control of the samples can be found in the Appendix [Appendix A].

### 2.2. Magnetic and geometric characterization of the Ti/Pt/Co sample

In a first step, to assess the magnetic properties of the Ti/Pt/Co sample, it was characterized with a vibrating-sample magnetometer (VSM, Quantum Design MPMS3). The hysteresis loops measured in fields perpendicular (⊥) or parallel to the sample surface (∥) (**Figure 1 a**) show that the sample exhibits a perpendicular magnetic anisotropy (PMA), i.e. an easy magnetization axis in the out-of-plane direction resulting in a perpendicular magnetization in the absence of external magnetic fields. The ratio of residual magnetization over saturation magnetization ($M_R/M_s$) is approximately 1. As can be seen in the zoomed in plot (**Figure 1 b**), for perpendicular fields the magnetization reversal from the homogeneously magnetized state starts at about 2 mT with an imminent drop in the magnetization, followed by a tail when approaching saturation. This shape of the hysteresis loop is typical for PMA samples with stripe domain structure [33] and can be attributed to the dipolar stabilization of stripe domains in low fields and their stepwise annihilation with increasing fields until full saturation is reached. The appearance of a stripe domain structure is confirmed by MFM measurements (s. inset in (**Figure 1 b**)). A self-correlation-based analysis [Appendix B] gives an average domain width of $< D^{MFM} > = 345\ nm$. The saturation magnetization $M_s$ is determined from the VSM measurements after correcting for sample shape effects [34] as $M_s = 201\ \text{kAm}^{-1}$. The uniaxial magnetic anisotropy constant $K_{u1} = 81\ \text{kJm}^{-3}$ is derived from the saturation magnetization field $H_{sat}$, which is extracted from the hard axis measurement [Appendix C] using the Stoner and Wohlfarth approximation [35] following:

$$K_{u1} = \frac{\mu_0 H_{Ku1} M_s}{2} \quad (5)$$

where the anisotropy field is given by [25];

$$H_{Ku1} = H_{sat} + M_s. \quad (6)$$

The observed perpendicular magnetic anisotropy (PMA) indicates a high quality of the Co-Pt interfaces since the PMA is induced by interface spin-orbit coupling which, in turn, is influenced by surface roughness [37]. Recent studies on a Ti/Pt/Co tri-layer material [38] showed besides asub-nm RMS roughness a Ti underlayer promoted textured fcc (111) growth, also enhancing the PMA [Appendix A].

AFM characterizations of the here used Ti/Pt/Co stack show an RMS roughness of 0.6 nm of the total stack, also further confirming the good structural properties. Such roughness values are neglectable for stray field simulations. From the QCM calibration we estimate a pessimistic upper limit for the stack thickness uncertainty of $d = 4$ nm, which is regarded in the stray field uncertainties calculated below [Appendices A and G]. For similar deposition rates and low sputtering powers, the thicknesses are confirmed [39,40].".

To further validate the useability of the Ti/Pt/Co sample as a stripe domain MFM reference sample, in the following we will determine its equilibrium zero field magnetic structure and magnetic stray field distribution by different means and compare the results to demonstrate the calculability of the Ti/Pt/Co sample. The different approaches and characterization routes that will be used are summarized in the validation flow diagram (**Figure 2**):

- The *domain pattern comparison* is used to prove that we understand the micro-magnetics of the Ti/PtCo material.
- The *tpc stray field comparison* serves the purpose to demonstrate that our reference sample is well understood and thus calculable and that different approaches (micromagnetic simulations, discrimination + forward calculation, qMFM) give the same magnetic stray field.
- The *IFW stray field comparison* will show that the Ti/Pt/Co sample, when actually used as a reference sample, gives correct quantitative stray field data in calibrated measurements, as validated by a comparison of Ti/Pt/Co-calibrated qMFM data on the Co/Pt sample with the results from discrimination and forward calculation.
- The *ICF comparison* finally will show that we do not merely get a proper quantitative analysis of "unknown" samples, but also a very good agreement of the $ICF$s and the thereof derived tip magnetic properties compared to calibrations with another reference sample.

We start with the topmost branch, the micromagnetic simulations, and progress with qMFM based on a calibration using a pre-existing Co/Pt multilayer reference sample [23,32] in the following depicted as 'ref'. We will refer to the different branches of the flow diagram where appropriate.

2.3. Micromagnetic Simulations of the Ti/Pt/Co sample's magnetization structure

The magnetization structure of the Ti/Pt/Co sample is modeled by micromagnetic simulations based on the Landau-Lifshitz-Gilbert (LLG) equation. The simulations are performed on an open-source GPU-accelerated micromagnetic simulation software MuMax³ [41] over a $1024 \times 1024 \times 20$ cell grid with a cell size of $5 \times 5 \times 3.8\ \text{nm}^3$ starting from a random magnetization distribution. The exchange stiffness $A_{ex}$ is slightly varied throughout the simulations within the range of $A_{ex}$values discussed in literature for similar magnetic multilayers ($5\ \text{pJm}^{-1} - 15\ \text{pJm}^{-1}$) [42,43]. Considering an optimum recovery of the experimentally observed domain width of $< D^{MFM} > = 345$ nm, a value for $A_{ex} = 6\ pJm^{-1}$ is derived. A long-range Ruderman-Kittel-Kasuya-Yosida (RKKY) exchange coupling was incorporated to the simulations which arises due to the Ti/Pt layers stacking. An effective RKKY exchange field $J_{RKKY}$ was implemented to the simulations by scaling the exchange coupling between each layer. The scaling factor is defined by $\Delta S = \frac{(J_{RKKY} \cdot \delta c_z)}{(2\langle A_{ex} \rangle)}$ where $\delta c_z$ and $\langle A_{ex} \rangle$ are the thickness of the single simulation cell and the average of $A_{ex}$ over the coupled layers [44]. Similar to what was done in the case of the exchange stiffness, $J_{RKKY}$ was varied throughout the simulations and the optimum value is found to be $J_{RKKY} = 0.07$ mJm$^{-2}$. The simulation results for optimized parameters are summarized in **Figure 3**.

The simulated sample reflects the stripe-like domain structure as the stable equilibrium magnetization pattern (**Figure 3a**), that was found by MFM. The magnetization of the sample was found to be homogeneous throughout the layer structure and independent of the thickness within the domains (**Figure 3c**), while the domain transitions areas show a slight layer dependence of the magnetization. Therefore, in **Figure 3a** a normalized averaged simulated z-component of the magnetization distribution $< m^{sim} >$ is shown. The average domain width of the simulation is found to be $< D^{sim} >= 370$ nm by using a self-correlation transform of the magnetization distribution, being slightly higher than the domain width found from MFM ($< D^{MFM} > = 345$ nm). While the average domain sizes found from the micromagnetic simulations and MFM differ slightly, the stripe domain pattern is well reproduced so that we consider the simulations as credible. The discrepancies can be attributed to simplifications in the simulations, e.g., neglecting grain structures, and to the finite simulation volume and discretization.

The good agreement justifies to determine an effective domain wall transition width from the simulations. **Figure 3b** shows a line-plot of the magnetization of the Ti/Pt/Co top layer sample along the line in the zoomed-in part of **Figure 3a**. The white dashed line shows the averaged angle of magnetization rotation. The domain wall width $\delta_{DW}$ using the definition of Lilley [45] can be calculated form the micromagnetic material parameters using:

$$\delta_{DW} = \pi \sqrt{\frac{A_{ex}}{K_{u1}}} \tag{7}$$

as $\delta_{DW}^{tpc} = 27$ nm, as indicated in **Figure 3c** with the dotted lines. Using $\delta_{DW}^{tpc}$, the simulation results thus can consistently be described with the standard 180° Bloch domain wall model [46]. Figure 3b shows the simulated domain transition together with $m_z = \tanh\left(\frac{x-x0}{\delta_{DW}}\right)$. The excellent agreement justifies to introduce domain walls based on a Bloch wall domain kernel [Appendix D] as discussed in [32].

Thus, all parameters required to calculate the reference effective magnetic charge density and stray field from the discriminated MFM image are now available. They are summarized in **Table 1**, together with the data of the Co/Pt reference sample.

**Table 1.** Sample parameters required for the calculation of the effective magnetic charge density of the Ti/Pt/Co sample and the Co/Pt reference sample

| | Ti/Pt/Co **multilayer Stack** | **Co/Pt stack** |
|---|---|---|
| Saturation Magnetization $M_s$ | 201 kA/m | 500 kA/m |
| Stack Thickness t | 20x3.8 nm | 100x1.3 nm |
| Domain Wall Width $\delta_{DW}^{tpc}$ | 27 nm | 16 nm |

2.4. Validation with qMFM and stray field simulations

To further validate the calculability of the Ti/Pt/Co sample, we characterize the sample with qMFM measurements based on a calibration with another reference sample (i) and compare the results with simulations based on the domain pattern determined from the MFM measurements using the material parameters found from the micromagnetic simulations and the VSM characterizations (ii) as well as with the results from micromagnetic simulations with predefined initial magnetization (iii). These comparisons are depicted in **Figure 2** as 'Domain Pattern Comparison' and 'TPC Stray Field Comparison', respectively.

(i) qMFM characterization of the Ti/Pt/Co sample

For a calibrated qMFM characterization, we used the pre-existing Co/Pt multilayer reference sample for the ICF calibration. It has a layer architecture of Ta(5 nm)/Pt(5 nm)/[Pt(0.9 nm)/Co(0.4 nm)]$_{100}$/Pt(2 nm). Similar to the Ti/Pt/Co sample, it shows a stripe domain pattern, at zero field, however with a lower average domain size $< D^{MFM,ref} > =$ 235 nm. The magnetic parameters of the sample are summarized in **Table 1**. The MFM measurements were performed with a Nanoscope IIIa with a Dimension head using a NT-MDT Low Moment MFM tip, following the procedure discussed in [32]. The measurement heights were $z^{ref} = 64$ nm and $z^{tpc} = 64$ nm for calibration and validation measurements, respectively, with a pixel size of $\delta_A = 10 \times 10$ nm$^2$ on a $512 \times 512$ spatial pixel grid. The quality factor $Q = 250$ was determined by fitting the resonance curve of the tip with a Lorentzian function. The full width of the resonance curve at 0.707 of the maximum was used as the $Q$ [32]. The cantilever stiffness c = 3 N/m was provided by the manufacturer. The $ICF^{ref}$ will be further discussed below and is shown in **Figure 6a**. The $TTF^{ref}$, i.e. the z-component of the stray field gradient $\mu_0 \frac{dH_z^{tip}}{dz}$ of the tip at the sample surface, calculated from the $ICF^{ref}$ using Eq. 1, is shown as the red line in **Figure 6c.** Before the calibrated measurement, i.e. the deconvolution as described in Eq. 3, the $\mu_0 \frac{dH_z^{tip}}{dz}$ distribution is circularly averaged around the center in order to eliminate artefacts arising from fast Fourier transforms (FFT).

In **Figure 4a,** an MFM image of Ti/Pt/Co sample taken with the calibrated system, exhibiting the expected stripe domain pattern, is shown. **Figure 4b** shows the stray field distribution of the Ti/Pt/Co sample $\mu_0 H_z^{tpc}$ at $z' = 64\,nm$ which is calculated from the MFM data by a deconvolution using the $TTF^{ref}$ yielding calibrated $\sigma_{eff}^{tpc}$ data. The stray field projected to the measurement is then calculated from these calibrated $\sigma_{eff}^{tpc}$ data using the following relation [13]:

$$\mu_0 H_z^{tpc}(\boldsymbol{k}, z') = \sigma_{eff}^{tpc}(\boldsymbol{k}) \cdot \frac{1}{2} e^{-\boldsymbol{k}z'}, \quad (8)$$

where $z' = z^{tpc} = 64$ nm. The qMFM measured stray field of the Ti/Pt/Co sample is found to be varying between $\pm 60$ mT.

(ii) MFM domain pattern-based simulations

Beyond using the MFM measurements to reveal calibrated stray field data, the MFM data allows an educated guess of the underlying domain pattern which then can be exploited to simulate the sample effective magnetization and stray field using the above determined material parameters. In this approach, the domain pattern is found from the magnetization configuration derived by the discrimination of the MFM data followed by a convolution with a domain wall kernel, as discussed above, resulting in the normalized magnetization distribution $m_z^{cal}(\boldsymbol{x})$ (**Figure 4c**). The resulting stray field distribution is again calculated from the effective surface charge density using Eq. 4 and Eq. 8 and is shown in **Figure 4d.**

(iii) Micromagnetic simulations

For better comparability with the MFM data, the micromagnetic simulations were repeated with the same material parameters as used for the simulation in **Figure 3a**. While a statistical initial magnetization is used in section 2.3 to prove the good agreement of the domain sizes of the simulation with the experiment, we here use the magnetization pattern $m_z^{cal}(\boldsymbol{x})$ derived from the MFM image as initial magnetization. The effective surface charge density and stray field of the simulated sample here is calculated using a multilayer approach that sums up the simulated magnetization distributions of all individual Ti/Pt/Co/Pt layers by:

$$\mu_0 H_z^{sim}(\boldsymbol{k}, z') = \frac{1}{2}\sum_i m_i^{sim}(\boldsymbol{k}) \cdot M_s^{tpc} \cdot \left(1 - e^{-\boldsymbol{k}\delta c_z}\right) \cdot e^{-\boldsymbol{k}z_i} = \frac{1}{2}\sigma_{eff}^{tcp} \cdot e^{-\boldsymbol{k}z_i} \quad (9)$$

where $\delta c_z = 0.5$ nm is the thickness of each Co layer, $m_i^{sim}$ is the relaxed magnetization distribution of $i^{th}$ layer and $z_i$ is the distance between the $i^{th}$ layer and the measurement height $z' = 64$ nm**. Figures 4e** and **f** show the averaged relaxed magnetization distribution $m_z^{sim}(\boldsymbol{x})$ and the resulting stray field, respectively. The simulation results well-reflect the characteristics of the experimentally observed domain structures.

In **Figure 4g** the results of all the approaches are compared in the form of lines plots through the calculated stray field distributions along the dotted lines in **Figures 4b, d** and **f**. The shaded bands show the uncertainties, calculated by a GUM (Guide to the expression of uncertainty in measurement [47]) conform approach propagating variances [48]. The same approach is used throughout the manuscript for uncertainty calculations. The ingoing variances are summarized in [Appendix E]. The magnetic field amplitudes of all three approaches agree very well. The discrepancies not covered by the uncertainty margins can be attributed to imperfections of the real sample like domains and pinning centers. This demonstrates that the material parameters of our Ti/Pt/Co sample are well understood. The micromagnetic simulations and their agreement with the qMFM measurements also validate the assumption of a homogeneous perpendicular magnetization in the domains all over the stack. Therefore, we consider the Ti/Pt/Co multilayer stack a well calculable reference material for qMFM.

## 2.5. Cross validation of the Co/Pt reference sample by Ti/Pt/Co calibrated qMFM

In a final validation step we switch the roles of the Ti/Pt/Co sample and the Co/Pt reference sample, i.e. we use the Ti/Pt/Co sample as the reference sample to calibrate the MFM setup, as it is the actual objective of this manuscript, and use the thus determined $ICF^{tpc}$ ("tpc Based ICF" **in Figure 2**) for a quantitative qMFM characterization of the pre-existing reference sample. These resulting stray field data are then compared to stray field data calculated from a discrimination of the MFM phase shift data using the known material parameters, allowing for the "IFW Stray field comparison" in **Figure 2**.

This analysis uses the same MFM measurement data as for the qMFM characterization of the Ti/Pt/Co sample discussed above and thus the same measurement parameters. Accordingly, the instruments $ICF^{tpc}$ (**Figure 6a**) is now derived from the Ti/Pt/Co sample measurement using a domain guess and a subsequent calculation of the reference $\sigma_{eff}^{tpc}$ by using the now validated Ti/Pt/Co materials parameters as listed in **Table 1.**

The thereof calculated $TTF^{tpc}$ is shown as the blue line in **Figure 6c**. For comparability, both $TTFs$ in the figure are calculated for the same distance from the tip apex of 64 nm. The $TTFs$, reflecting the tip's magnetic properties which thus ideally should be independent of the used reference sample, show very good agreement, the small discrepancy might be attributed to the differences in the Fourier spectra of the two reference samples.

The quantitative $[Pt/Co]_{100}$ sample stray field data that result from applying the $TTF^{tpc}$ to the MFM phase shift data (**Figure 5a**) are shown in **Figure 5b**. **Figure 5c** and **d** show the magnetization distribution from the discrimination of the phase shift data and the thereof calculated stray field data, respectively. Line plots of the both stray field data distributions taken along the dashed lines and plotted together with their uncertainty bands are compared in **Figure 5f**. The Ti/Pt/Co -calibrated qMFM data show excellent agreement with the simulations mostly within the uncertainty margins. Again, small discrepancies can be explained by imperfections of the real sample not included in the uncertainty calculations.

### 2.6. Feature size spectra

While the above discussion demonstrates the usability of the Ti/Pt/Co stack from the standpoint of calculability and field range, the application range of the Ti/Pt/Co stack for qMFM based measurements of sample under test (SUT) must also be discussed in terms of the accessible feature size spectrum since magnetic features, i.e. characteristic magnetic structures of the sample under test, on length scales not covered by the reference material are suppressed in the calibrated measurement. The covered feature size of the reference sample can be quantified in the form of the Fourier spectrum of the sample's effective charge distribution as a function of spatial frequency (**Figure 6d**). However, the spectral spectrum accessible after a calibration also depends on the used tip and the respective MFM system, since the tip's magnetic stray field gradient distribution enters into the systems sensitivity and the MFM system's detection system determines the measurement noise floor. A detailed discussion can be found in [Appendix F]. **Figure 6d** features the Fourier spectra of both, the Ti/Pt/Co and the Co/Pt sample. Both reference samples show a significant overlap of their spectra, retrospectively justifying the qMFM characterization of the Ti/Pt/Co stack, even though the dominant components in the spectra of both samples are not identical. The lower amplitude of the Ti/Pt/Co spectrum is due to its lower saturation magnetization and lower thickness. The characteristic structure sizes accessible after a calibration for a Gaussian white noise with 0.2° standard deviation are found as (149 nm-5.12 μm) for the Ti/Pt/Co sample and (124 nm-5.12 μm) for the Co/Pt sample (indicated as dotted lines in **Figure 6d)**. A lower noise floor (as e.g. achievable in vacuum MFM) with 0.02° standard deviation would allow access to smaller structures sizes (112 nm -5.12 μm) for Ti/Pt/Co and (99 nm -5.12 μm) for Co/Pt.

The observed accessible structure size range thus renders the Ti/Pt/Co sample suitable for qMFM calibrations for quantitative measurements on magnetic micro- and nanostructures like stripe and bubble domains, skyrmions and antiferromagnetic domains.

## 3. Conclusion

The Ti/Pt/Co sample has been proven to be a suitable reference sample for qMFM calibrations, covering features from the 10 μm to 100 nm range and thus applicable to the quantitative characterization of relevant micromagnetic materials. For the first time, a systematic step-by-step reference sample validation process was established and pursued. The Ti/Pt/Co sample has a very low surface roughness and well-defined interfaces. The magnetic properties of the sample were defined by using micromagnetic simulations and macroscopically measured magnetic parameters. The magnetic structure found from the simulations was validated by qMFM measurements. The maximum stray field of the sample at 64 nm was found around $\pm 60\ mT$ where the reference sample used in the pass has a maximum field strength of more than $\pm 130\ mT$. The low field strength makes the Ti/Pt/Co sample a good candidate to be used as a reference sample for the calibration of low moment tips. The Ti/Pt/Co sample was successfully used as a reference sample for a qMFM stray field measurement of the pre-existing reference sample. The highly reproducible fabrication process guarantees a high availability of the reference material. The sample's inherent stripe domain pattern provides a broad spectrum of Fourier components with sufficiently high amplitudes. While artificially patterned samples might give higher flexibility in controlling the FFT spectra, patterning attributed artefacts make the effective magnetization and topography of patterned samples hard to control. Adapting the Ti layer thickness and the repetitions of the Ti/Pt/Co building block of the stack, exploiting the high stability of the deposition process, may, in future work, allow to further adjust the sample stray field amplitude and the width of the domain pattern, thus opening a path towards a fabrication of reference samples with properties tailored for specific applications.

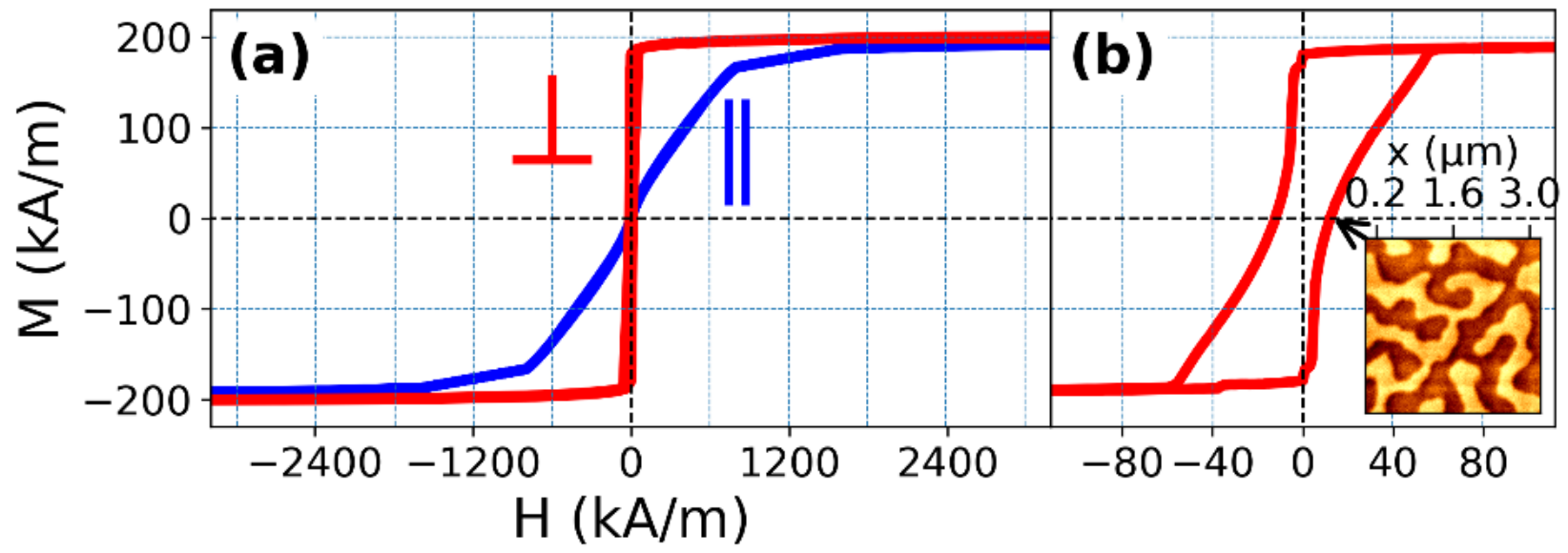

**Figure 1.** M-H hysteresis loop of the Ti/Pt/Co sample. (a) the easy axis and hard axis hysteresis loops recorded by external field applied in plane, parallel to the sample surface (∥, blue) and out of plane, perpendicular to the sample surface (⊥, red). Measurements were performed by using VSM at room temperature (295K). (b) zoomed-in plot of the out-of-plane measurement shown in (a). The inset in (b) shows an MFM image of the sample.

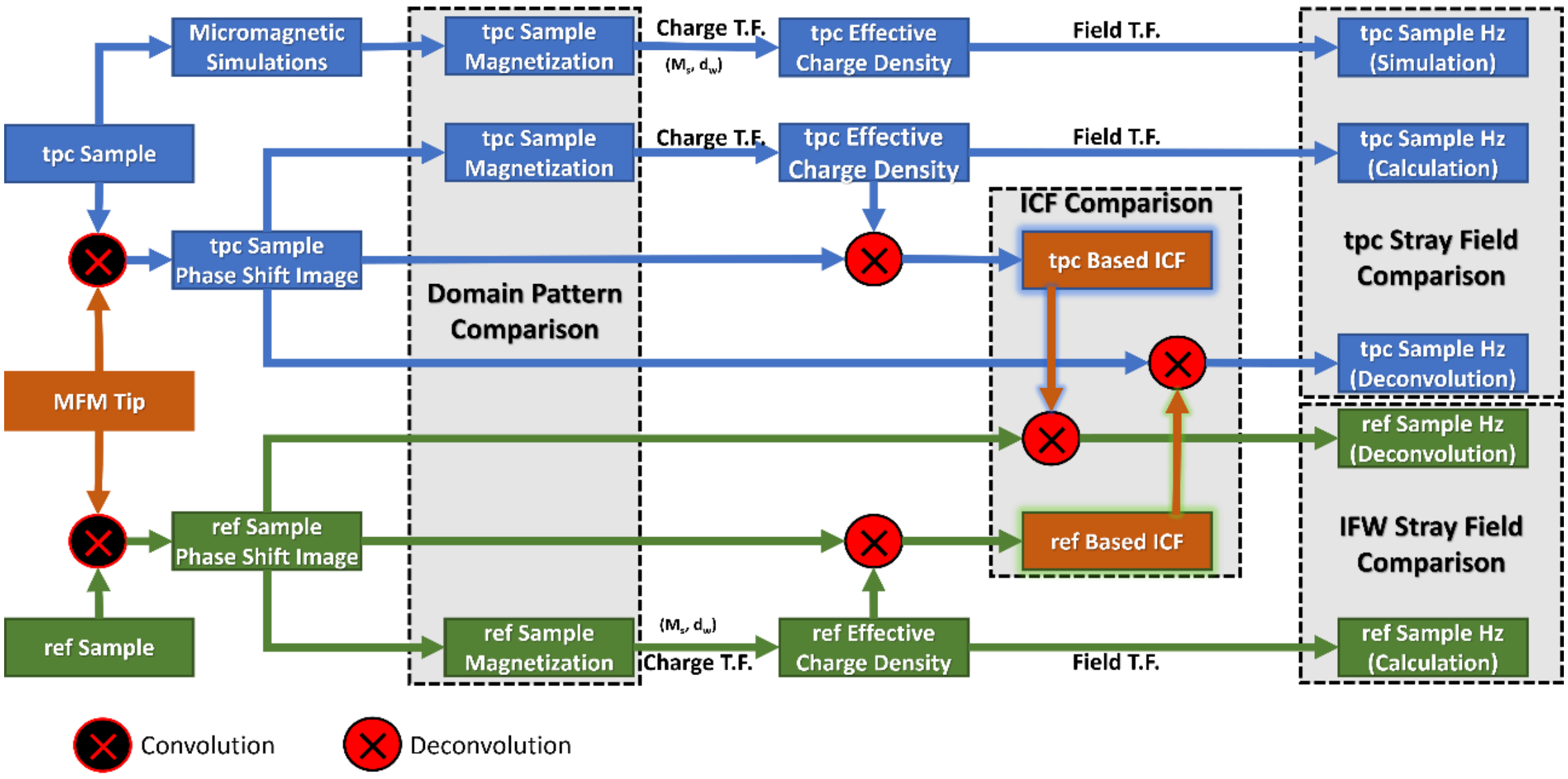

Figure 2. Flowchart of the validation process. The flowchart shows the different simulation and measurement steps used to validate the Ti/Pt/Co sample's micromagnetic parameters. The comparisons that were performed based on the measurement results are marked as grey shaded boxes.

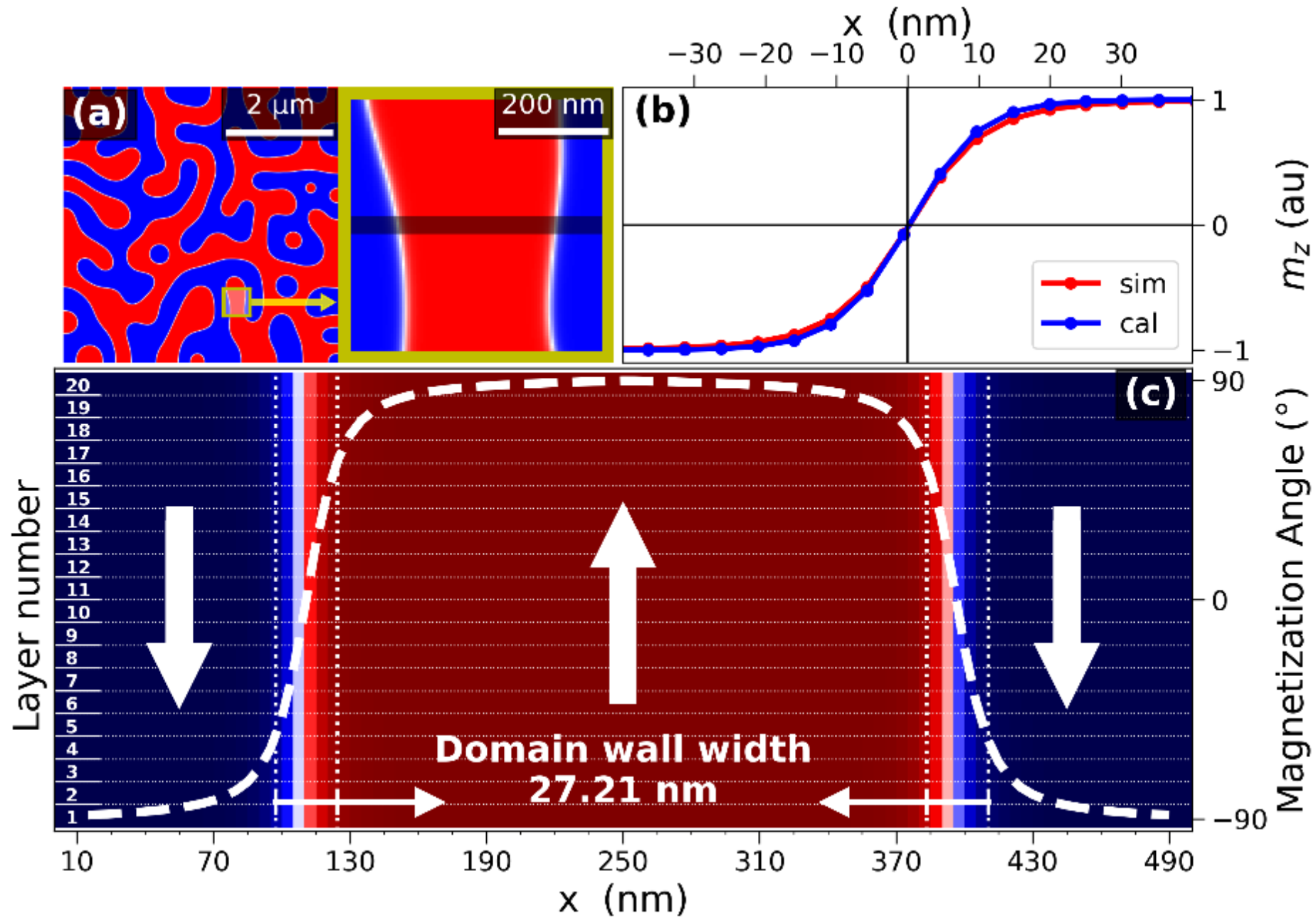

**Figure 3.** Results of the MuMax3 micromagnetic simulations of the Ti/Pt/Co sample: The relaxed magnetization pattern averaged over the stack thickness together with a zoomed in view (a). perpendicular magnetization component of the top layer in a transition between two domains together with a calculated transition using the standard Bloch wall model (b). Cross section of a cutout of the magnetization of the Ti/Pt/Co sample showing all 20 layers (c). The magnetization in the domains is homogeneous and independent on the layer number. The overall magnetization of the domains is depicted by the arrows. The dashed line shows the angle of the magnetization that follows a Bloch like course of the domain wall transition. The dotted lines mark the domain wall width calculated using the Lilley formula.

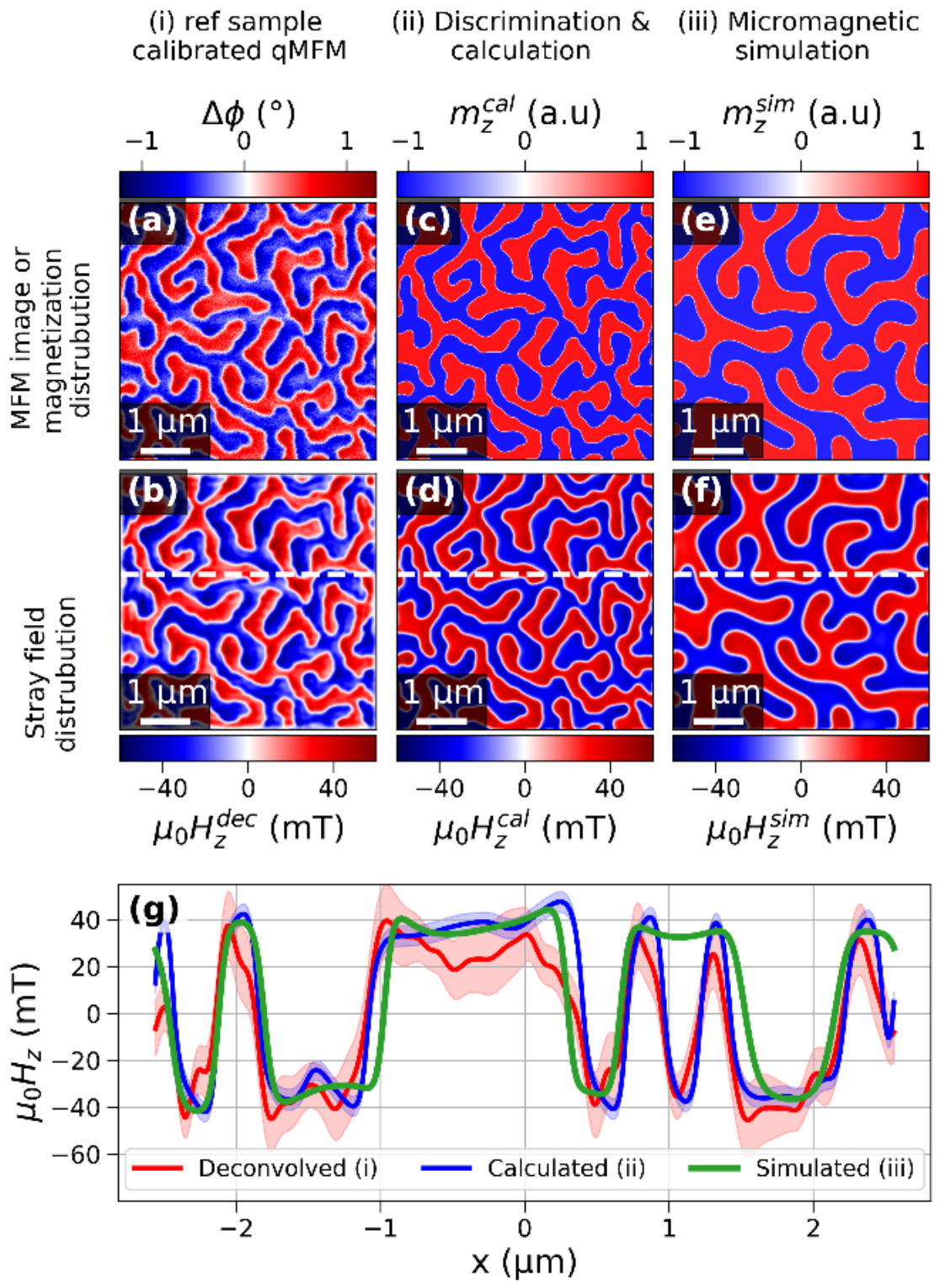

**Figure 4.** Comparison of simulated and experimental Ti/Pt/Co sample data. (a) measured MFM phase shift data and (b) perpendicular stray field components $B_z$ data calculated using the calibrated qMFM CoPt sample. (c) z component of the magnetization calculated from a discrimination of the MFM phase shift data from (a) and (d) the thereof calculated perpendicular stray field components $B_z$ data using the Ti/Pt/Co micromagnetic material parameters. (e) z-component of the magnetization from the micromagnetic simulation and (f) the perpendicular stray field components $B_z$ data calculated thereof using a layer by layer approach. (g) the stray field data with uncertainty bands from the data marked by the dashed lines in the stray field images in (b), (d) and (f).

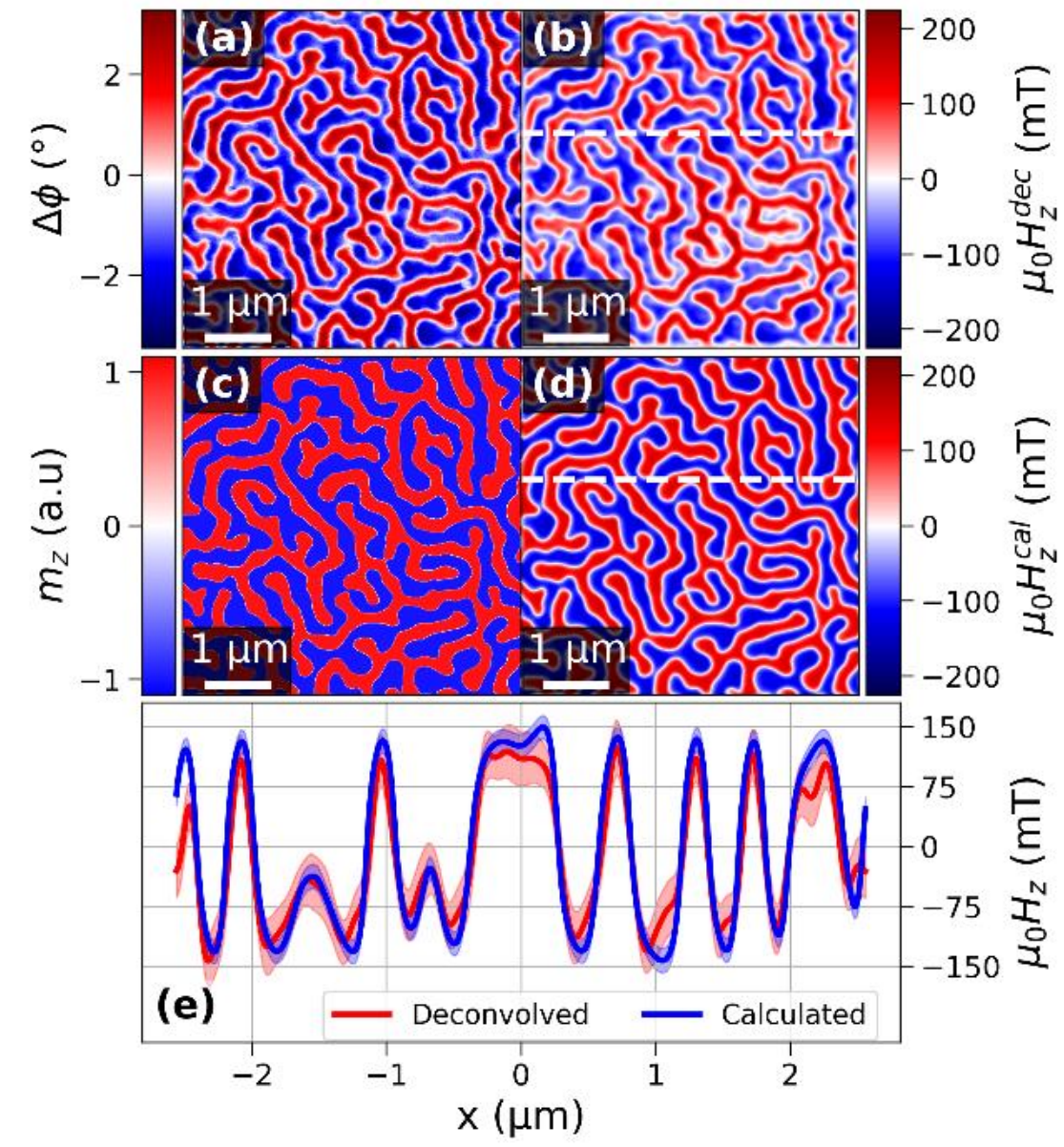

**Figure 5.** (a) MFM phase shift data of the $[Pt/Co]_{100}$ sample and (b) quantitative perpendicular stray field components $B_z$ data calculated thereof using Ti/Pt/Co calibrated qMFM; (c) sample magnetization pattern from discrimination of the phase shift data followed by a convolution with a domain wall kernel and (d) perpendicular stray field components $B_z$ data calculated thereof by forward simulation using the known micromagnetic material parameters. (e) shows plot lines of the perpendicular stray field components $B_z$ taken along the dashed lines the stray field images together with uncertainty bands.

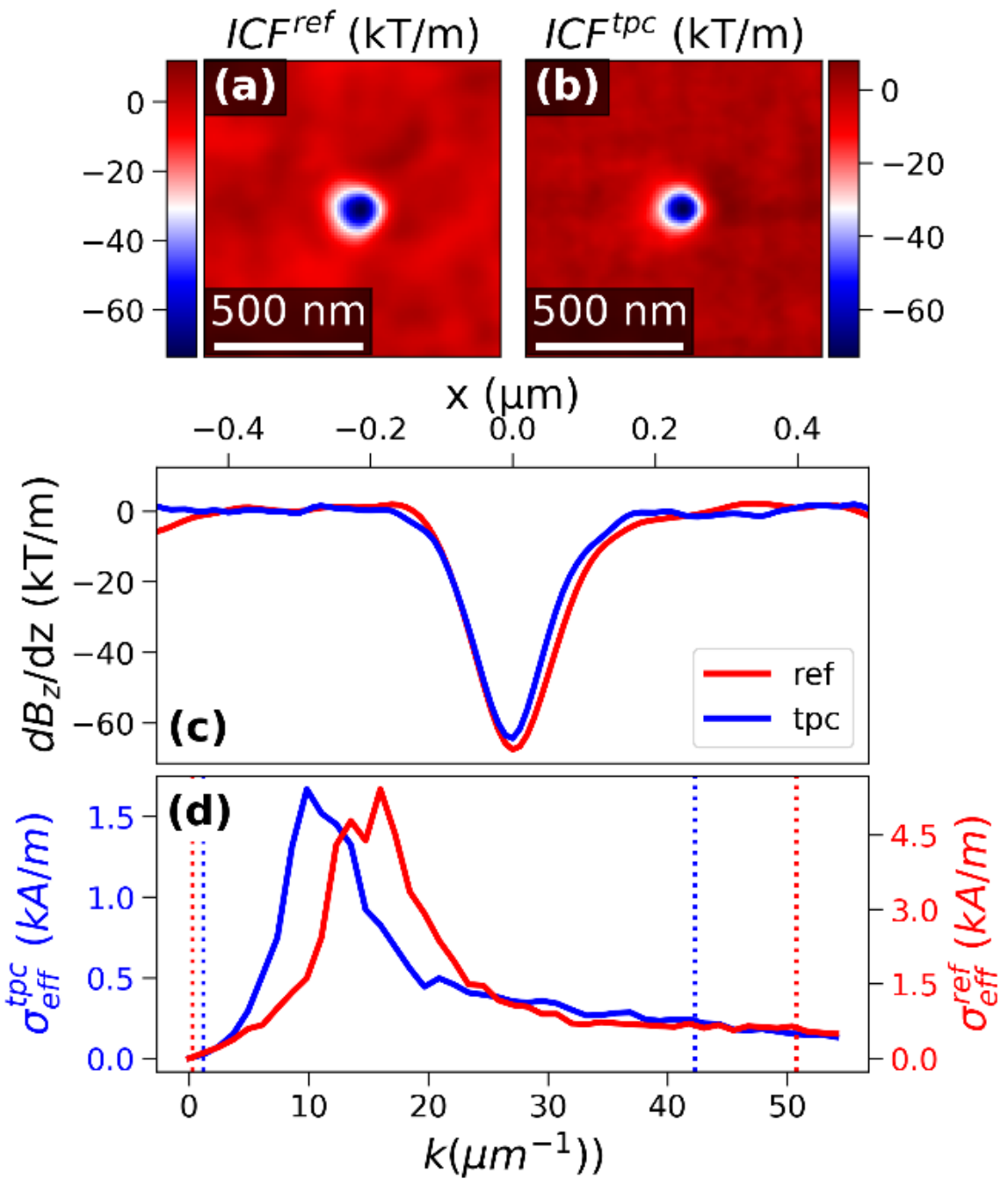

**Figure 6.** Comparison of $ICFs$ and reference sample spectra: (a) and (b) show the $ICFs$ calculated from calibration measurements using the Co/Pt ($ICF^{ref}$) and the Ti/Pt/Co ($ICF^{tpc}$) reference sample, respectively; (c) plot lines through the maxima of the $TTFs$ for both calibrations, $TTF^{ref}$ (red) and $TTF^{tpc}$ (blue) for a distance of 64 nm from the tip apex. (d) shows plotlines through the Fourier spectra of the effective surface charge density of the Co/Pt reference sample (red) and Ti/Pt/Co sample. Dotted line marks the area of k-values accessible after a calibration with the respective reference sample (see text).

**Acknowledgments:** We thank Volker Neu from Leibniz IFW Dresden for providing the Co/Pt reference sample and material parameter data.

**Author Contributions:** B.S. and I.O. deposited and provided the TPC multilayer stack. A.F.S. and S.S. provided VSM and MFM data. B.S. and F.G.S. carried out the micromagnetic simulations. B.S. and S.S. discussed and performed the numerical analysis. O.Ö. and H.W.S. initiated and coordinated the study. B.S. and S.S. wrote the manuscript.

All authors have read and agreed to the published version of the manuscript

**Funding:** This research was funded by the Scientific and Technological Research Council of Turkey (119M287) and by the European Metrology Research Programme (EMRP) and EMRP participating countries under the European Metrology Programme for Innovation and Research (EMPIR) Project No. 17FUN08-TOPS Metrology for topological spin structures. B.S. acknowledges support from the Scientific and Technological Research Council of Turkey, International Research Fellowship Program for PhD Students (2214-A/1059B141800226).

**Data Availability Statement: T**he data that support the findings of this study are available on request from the corresponding author [B.S.].

**Conflicts of Interest:** The authors declare no conflict of interest. The funders had no role in the design of the study; in the collection, analyses, or interpretation of data; in the writing of the manuscript, or in the decision to publish the results.

**Appendix A: XPS study and calibration of deposition**

Deposition rates of the magnetron sputter depositions are calibrated by using the XPS technique. Rates are calculated in means of intensity change of the XPS signal over deposition time. In this calibration, the change of the substrate's XPS signal was analyzed for specific deposition parameters (i.e., Ar flow, Target-Substrate distance, Sputtering Power etc..).

The substrates are Ag foil for Pt and Ti, Au foil for Co calibrations. Substrates are mechanically polished beforehand. After loading into vacuum, the sequential of Ar plasma etching and XPS scan was used ensuring the purity of the substrates. For each material, XPS main peak intensities of the substrate (Au4f for Co, Ag3d for Pt and Ti) was measured after each definite sputtering duration as well as pristine state of the substrate (Fig. A1a-c). Thickness calculation of the deposited film at each sputtering step was done by using the well-known modified Beer-Lambert equation [49];

$$I_i = I_0 \cdot e^{\lambda/d_i} \tag{A1}$$

where $I_0$ and $I_i$ are the peak intensity of the relevant substrate and energy level before and after the $i^{th}$ deposition, $\lambda(E)$ is the inelastic mean free path of the electron emitted from the substrate's relevant energy level while passing through the deposited material and $d_i$ is the thickness of the material grown on the substrate at $i^{th}$ deposition. The inelastic mean free path values obtained from NIST's database, Tanuma model were $\lambda_{Pt}(1118.47\ eV) = 14.17$ Å , $\lambda_{Ti}(1118.47\ eV) = 25.27$ Å and $\lambda_{Co}(1169.55\ eV) = 17.38$ Å [50]. In order to calculate the deposition rate, thickness dependence on the deposition time was linear fitted (Fig. A1d). These rates were used to calibrate the quartz crystal microbalance (QCM) thickness monitor (Maxtech Inc., 6kHZ reference crystal). QCM placed next to the sample holder and used to monitor the thickness during the deposition. In the study reported by Melek et. al, the effect of 0.3 nm Ti layer on magnetic anisotropy was investigated in a similar sample architecture and found out that the Ti promote the Pt to be grow in a fiber texture structure. In the same study it was reported that in the presence of Ti underlayer the surface morphology changed dramatically and the surface/interface roughness decreased [Melek 2021].

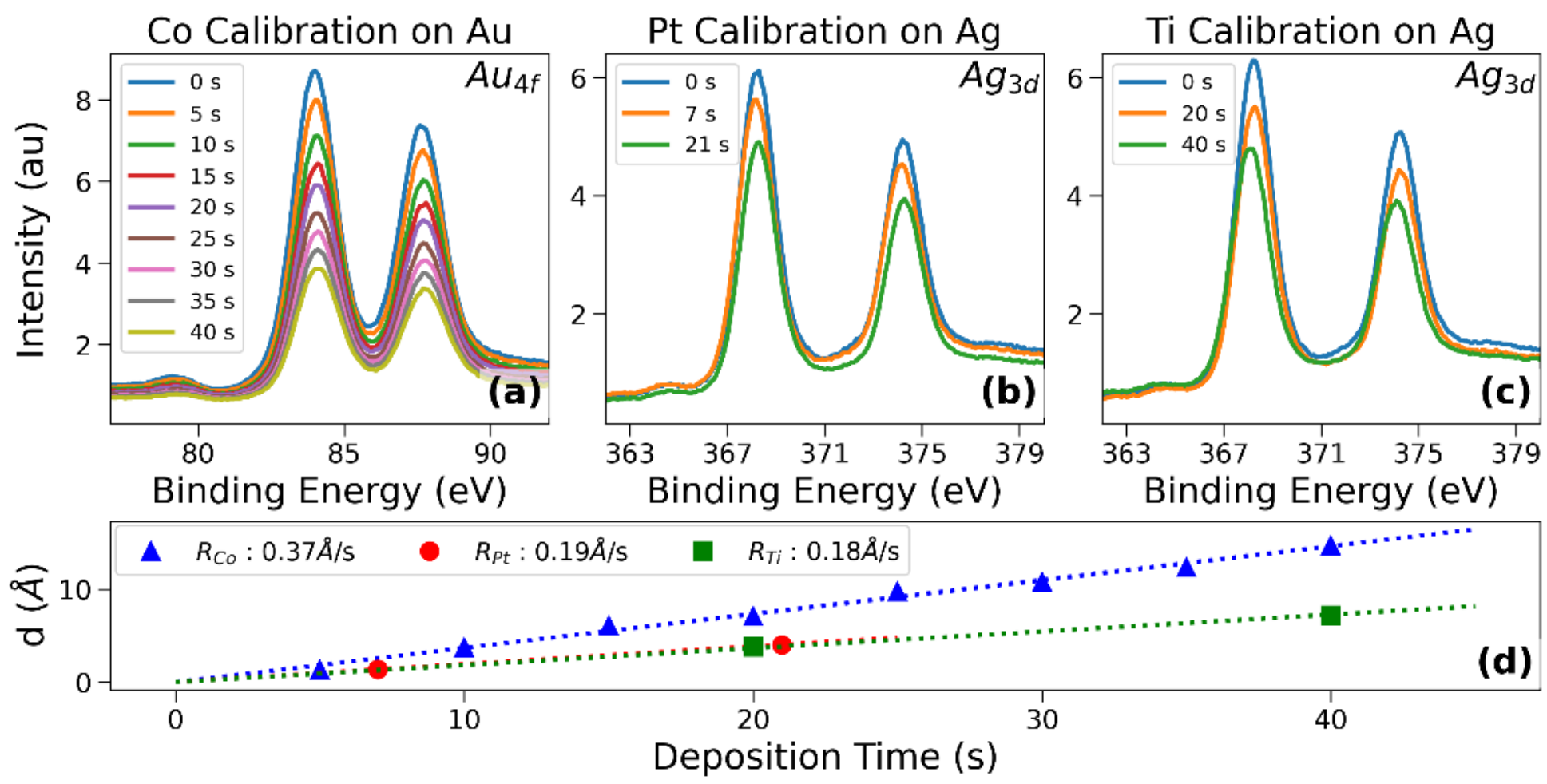

**Figure A1.** X-ray photoemission spectra of Au4f and Ag3d used for calibration of Co (a), Pt (b) and Ti (c) depositions. Thickness – deposition time graph and deposition rates for Co, Pt and Ti targets (d).

An XPS spectrum of the samples is used to the ensure that the sample fabrication was as it was desired. XPS is a highly surface sensitive characterization technique and could detect any change in the electronical or morphological change in the films. The change in the interfaces would also change the cross-section of photoemission and would be detected in the spectrum.

In Figure A2a the wide range survey spectrum of the sample recorded after the deposition of the first Co layer is shown. Survey spectrum is the recorded with an energy step of $\Delta E = 1\ eV$. High-resolution windows of each element in the survey spectrum are recorded with a resolution of $\Delta E = 0.1\ eV$. Figure A2b-g shows the recorded high-resolution spectra of deposited platinum 4f, titanium 2p and cobalt 2p levels, substrate's silicon and oxygen 2s levels and also zero-binding energy region (Fermi region). These XPS spectra for repeated tpc block provide a "fingerprint" that ensures the high accuracy of the reproducibility.

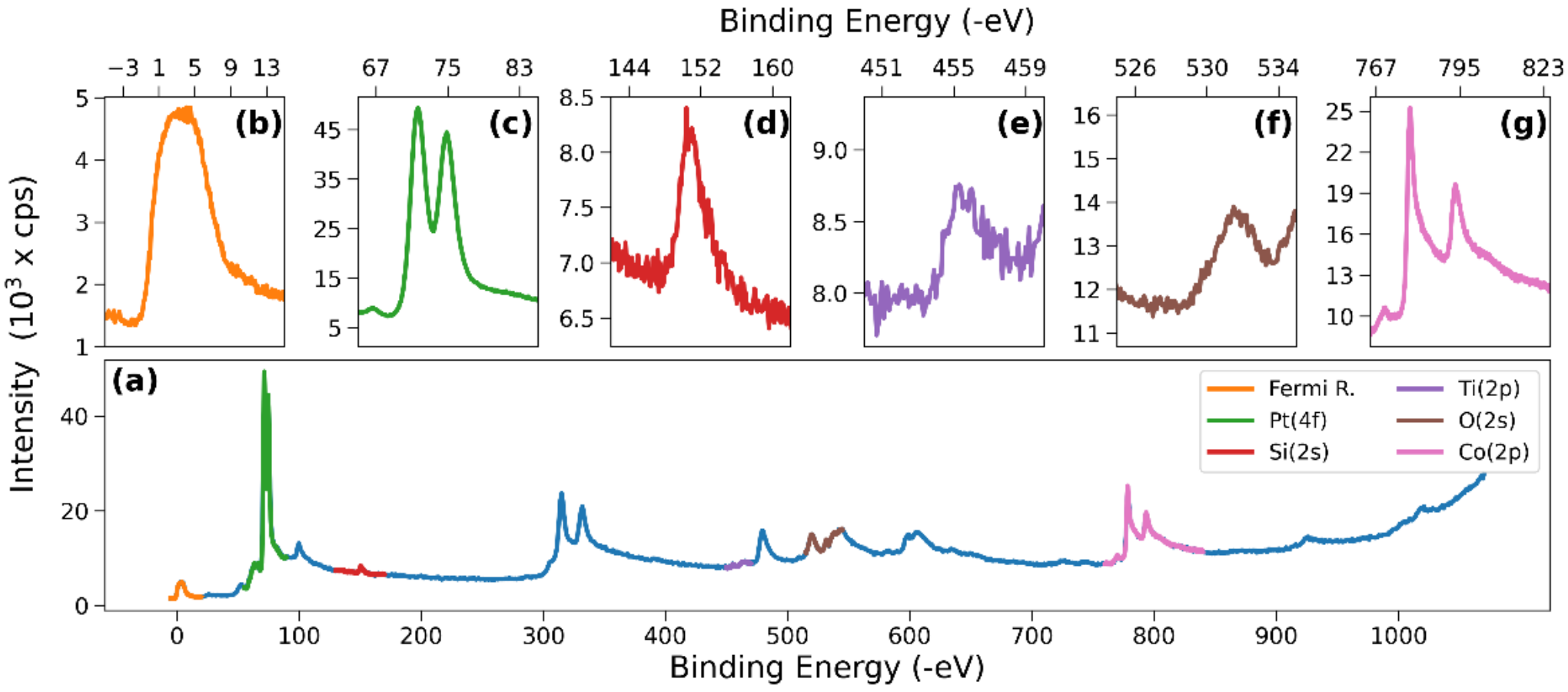

**Figure A2.** X-ray photoemission spectra of tpc sample recorded from the top of first Co layer. The survey spectrum is used to detect any contamination on the sample and check the general structure of the sample (a). High resolution windows of each layers and substrate is used to check the consistency of the photoemission cross section and the change in the chemical state of any layer (or substrate) (b-g).

## Appendix B: A self-correlation-based analysis of domain wall widths

The average domain size of a magnetization pattern is calculated from a self-correlation transform of the MFM data. The distance between the first and the second maximum gives twice the average domain size [51]. Plot lines through the self-correlation transforms for the Co/Pt refence sample, the tpc reference sample and the simulated MFM image of the micromagnetic simulation results are shown in **Figure A3**. The MFM image of the simulated magnetization pattern was calculated by first determining its effective surface charge density followed by a convolution with the $ICF$ as determined by a calibration measurement with the Co/Pt reference sample.

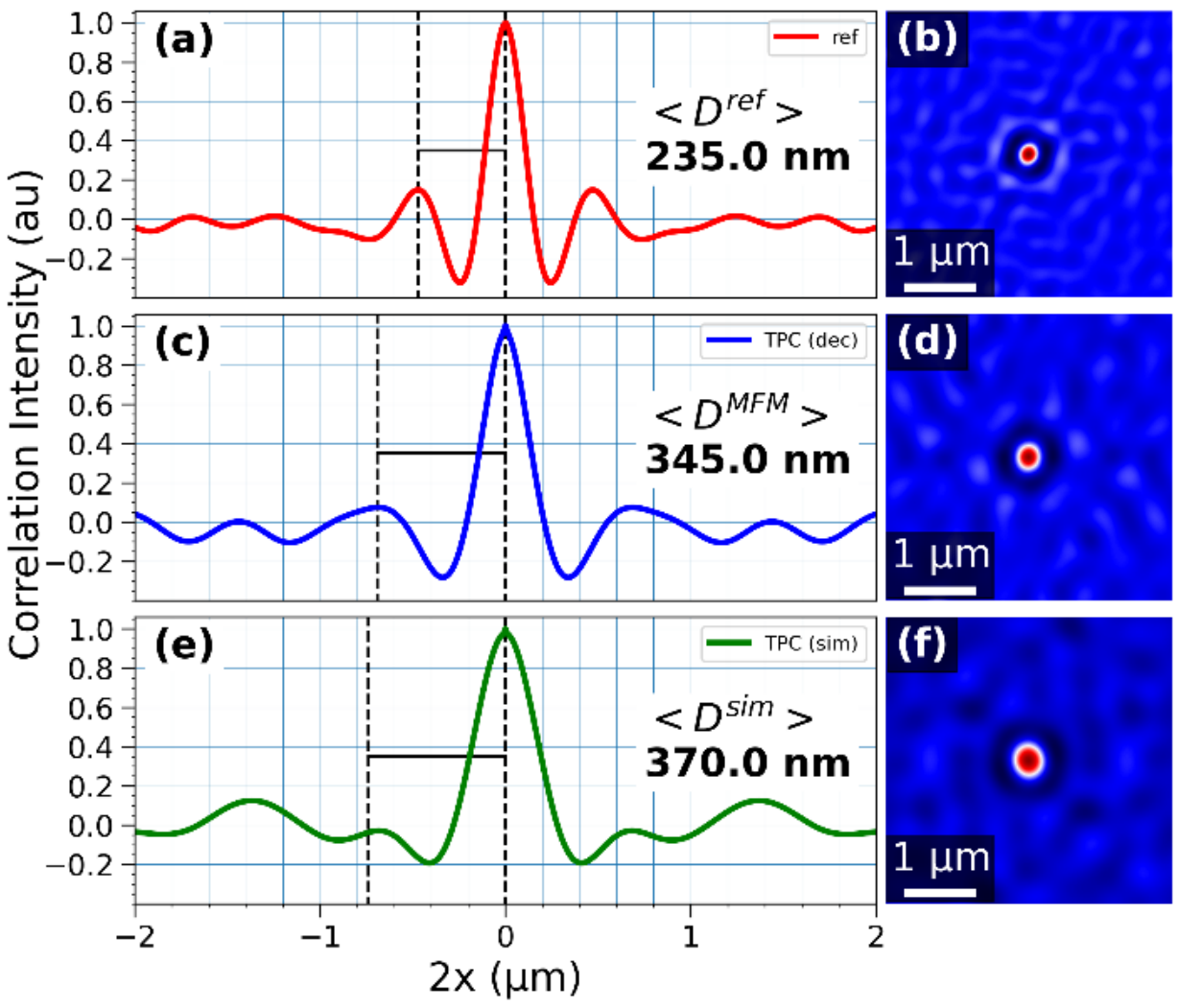

**Figure A3**. Plot lines through (a,c,e) the self-correlation transforms for the Co/Pt reference sample (ref) (b), the tpc reference sample (TPC (dec)) (d) and the simulated MFM image of the micromagnetic simulation results (TPC (sim)) (f).

## Appendix C: Determination of the Ti/Pt/Co sample's uniaxial anisotropy constant $K_u$ from the VSM data

To calculate the anisotropy constants from the VSM measurements, the hard axis in- plane M-H data are analyzed, following the Stoner Wohlfarth model. Thereto, the low-field linear part of the M-H curves is linearly fitted (**Figure S4a**) and the fitted function is extrapolated to find the intersection with the $M = M_{sat}$ dotted horizontal line. The corresponding $\mu_0 H$ value gives $\mu_0 H_{sat} = 0.6\ T$ (**Figure S4b**).

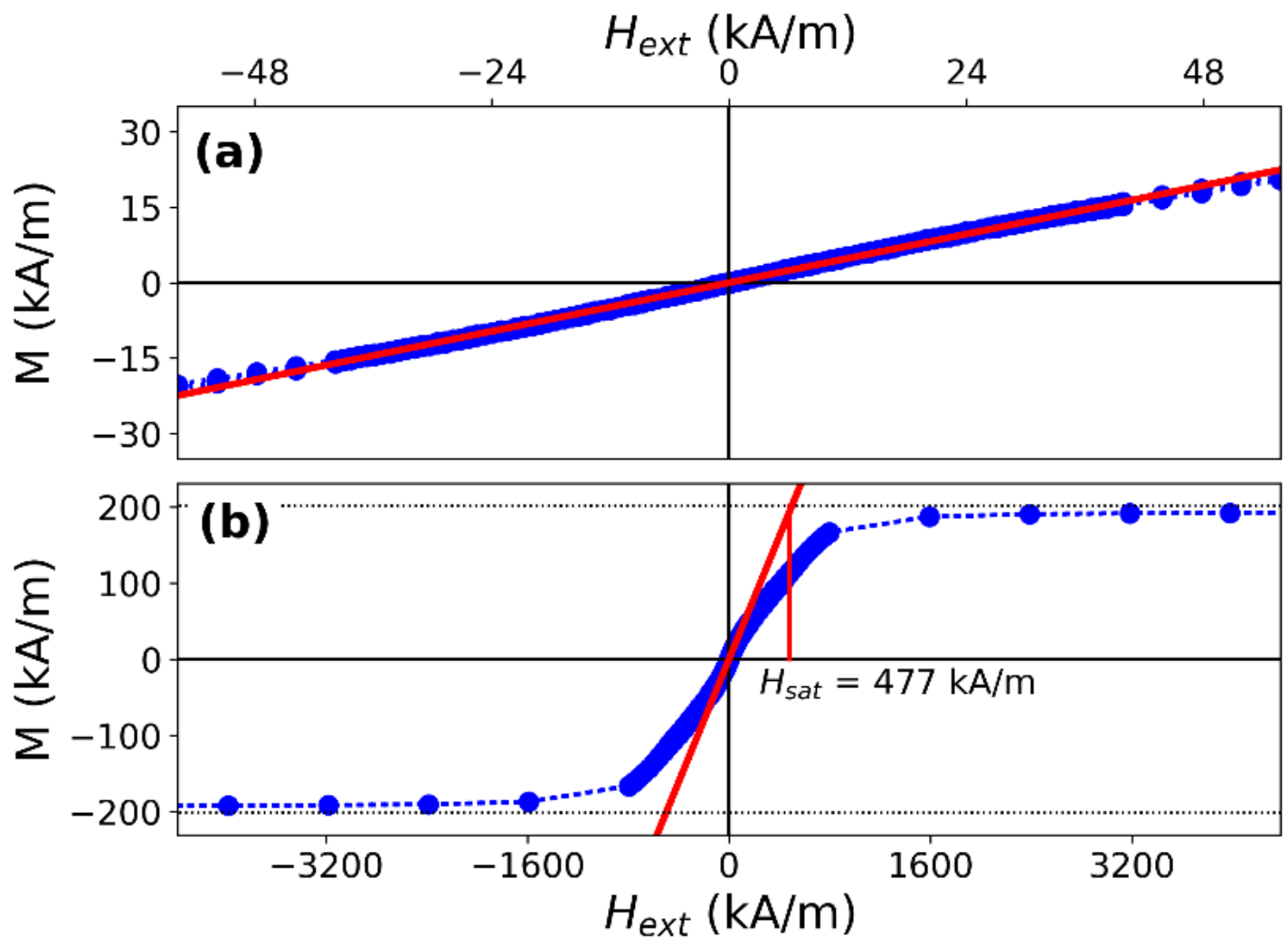

**Figure A4**. VSM measured M-H curves of the Ti/Pt/Co reference sample. (a) shows a zoomed-in version of (b). The red line shows the fit to the low-field linear part of the curve. (b) also shows the construction of the $\mu_0 H_{sat}$ value from the intersection of the fit with $M = M_{sat}$.

**Appendix D: Domain wall kernel**

Domain wall transitions are introduced to the binary magnetization patterns from discrimination by a convolution with a domain wall kernel. The kernel used to introduce tanh-like domain wall transition following the standard 180° Bloch wall model is;

$$f(x,y) = sech^2\left(\frac{\pi\sqrt{x^2+x^2}}{\delta_{DW}}\right) \quad \text{(A2)}$$

$\delta_{DW}$ is the domain wall width of the 180° Bloch domain wall.

**Appendix E: Uncertainties used in uncertainty calculations**

Uncertainty bands are calculated by a propagation of uncertainties following a GUM conform approach [47]. The relative uncertainties used are summarize d in **Table A1:**

**Table A1**. Uncertainties used for the calculation of the uncertainty bands.

| Parameter | uncertainty |
|---|---|
| MFM phase shift $\Delta\boldsymbol{\varphi}$ | $u_\Delta\varphi = 0.2°$ |
| regularization parameter $\boldsymbol{\alpha}$ | $u_\alpha$ : 1% |
| stack thickness tpc sample $\boldsymbol{d^{tpc}}$ | $u_d$ = 2 nm |
| stack thickness ref sample $\boldsymbol{d^{ref}}$ | $u_d$ = 4 nm |
| saturation magnetization tpc sample $\boldsymbol{M_S^{ref}}$ | $u_M_S^{ref}$: 6% |
| saturation magnetization Co/Pt sample $\boldsymbol{M_S^{tpc}}$ | $u_M_S^{tpc}$: 6% |
| measurement height $\boldsymbol{h}$ | $u_h$ : 10% |

**Appendix F: Estimation of accessible spatial frequency range**

The spatial frequency range accessible for qMFM after a calibration with the Ti/Pt/Co is not solely a property of the reference sample's effective charge density, $\sigma_{eff}(\boldsymbol{k})$, but rather depends on the combined contributions of the reference sample, the MFM tip and the used MFM system. This follows from the fact that, the tip stray field gradient distribution determines the MFM sensitivity following (the lever correction is neglected):

$$\Delta\Phi(\boldsymbol{k},z') = -\,\sigma_{eff}^{*}(\boldsymbol{k}) \cdot \frac{Q}{c} \cdot \frac{dB_z^{tip}}{dz}$$

During the calibration measurement, the MFM detection system determines the phase shift data noise level, which in turn determines the sensitivity of the system. To estimate the spatial frequency range accessible to a calibration, we employ a generic tip with a stray field gradient distribution modeled as a Gaussian function (see inset in **Figure A5a**) with typical values as found for real-world tips (Amplitude 60kT/m, $\text{sigma}_x$= $\text{sigma}_x$ =50 nm) .

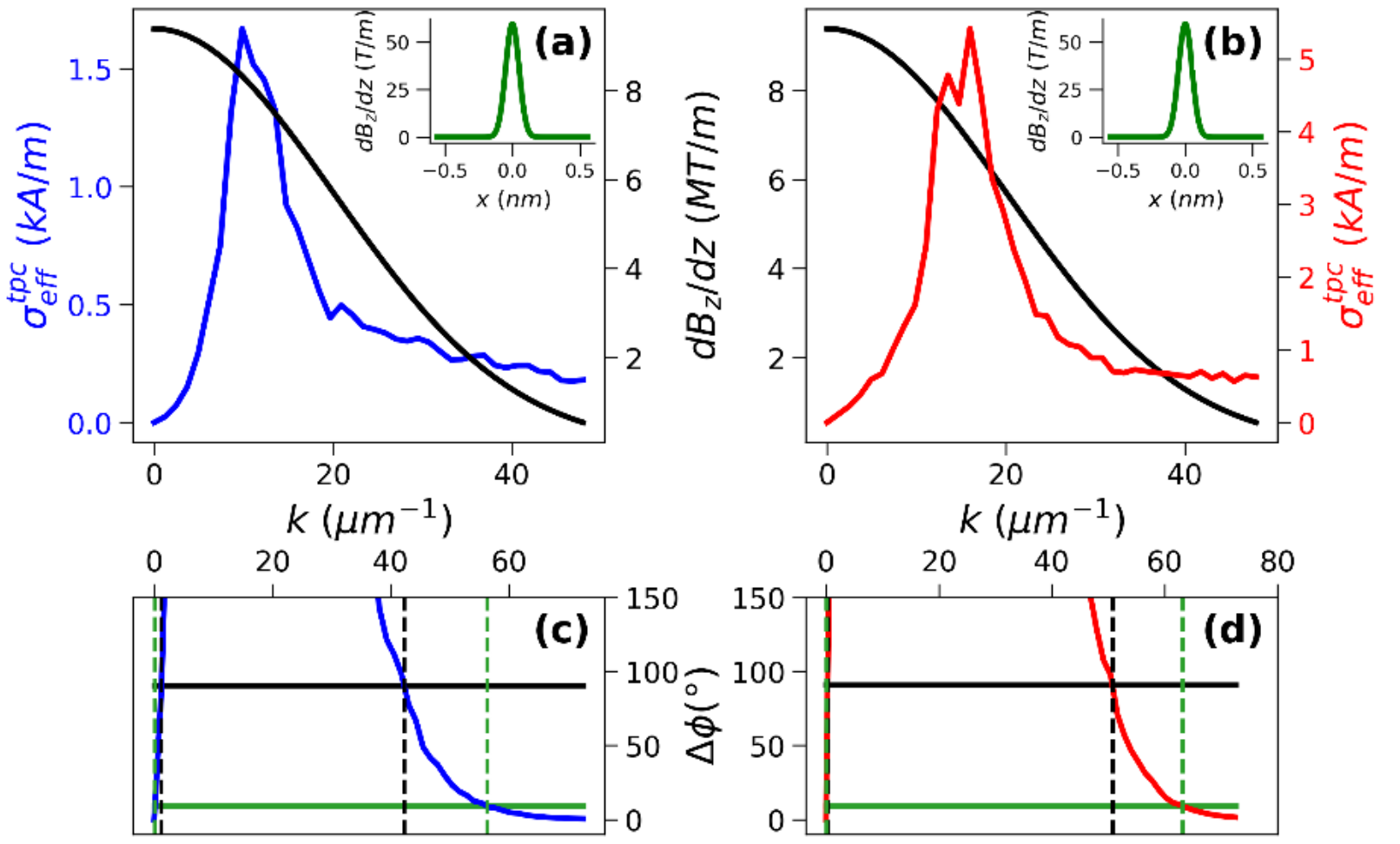

**Figure A5:** Estimation of accessible wave vector range after a calibration; tip mediated sensitivity $\frac{dB_z}{dz}$ and circularly averaged sample $\sigma_{eff}$ distribution in Fourier space for the Ti/Pt/Co (a) and the Co/Pt sample. The inset shows the generic tip's $\frac{dB_z}{dz}$ in real space. (c) and (d) show the phase shift distribution of a simulated reference sample measurement using the generic tip for the Ti/Pt/co and the Co/Pt sample, respectively. The horizontal lines show the noise floor for white Gaussian noise with 0.2° (black) and 0.02° (green) standard deviation.

The tip mediated sensitivity $\frac{dB_z}{dz}$ in Fourier space is plotted in **Figure A5a** together with the circularly averaged sample's $\sigma_{eff}$. A cut-out off the cross-section of the resulting phase shift Fourier spectrum $\Delta\Phi(\boldsymbol{k})$ of the Ti/Pt/Co sample (using $Q = 250$, c= $3N/m$) is plotted in **Figure A5c.** Additionally, the plot shows noise levels calculated for a Gaussian white noise with standard deviations 0.2° (typical for ambient conditions MFM, black horizontal line) and 0.02° (e.g. vacuum MFM, green horizontal line), respectively. The interception of the phase shift spectrum with the noise levels defines the low ($k_{low}$ ) and high ($k_{high}$) wave-vector cut-off frequencies. **Figures 5b** and **c** shows the analog analysis for the CoPt sample. In all cases, the lower cut off frequencies are limited by the sample size. **Table A2** summarizes the such derived k-vector and corresponding wavelength data calculated ac $\lambda = \frac{2\pi}{k}$, i.e. the range of characteristic structure sizes detectable with the MFM after the calibration with the respective reference sample.

**Table A2:** Cut-off wavevector and corresponding wavelength data for the Ti/PT/Co and the Co/Pt sample for two different noise levels.

| $\Delta\phi$ | Low Cut-off | | High Cut-off | |
|---|---|---|---|---|
| | **Frequency** | **Wavelength** | **Frequency** | **Wavelength** |
| **Ti/Pt/Co multilayer Stack (tpc)** | | | | |
| 0.02° | <1.22 μm$^{-1}$ | >5.12 μm | 42.256 μm$^{-1}$ | 149 nm |
| 0.2° | <1.22 μm$^{-1}$ | >5.12 μm | 56.295 μm$^{-1}$ | 112 nm |
| **Co/Pt Stack (ref)** | | | | |
| 0.02° | <1.22 μm$^{-1}$ | >5.12 μm | 50.726 μm$^{-1}$ | 124 nm |
| 0.2° | <1.22 μm$^{-1}$ | >5.12 μm | 63.231 μm$^{-1}$ | 99 nm |

### Appendix G: Estimation of the Ti/Pt/Co sample surface roughness

To characterize the surface roughness, the Ti/Pt/Co sample was characterized with atomic force microscopy in a Park Nx-Hivac system using a PPP-NCHR tip.

**Figure A6a** shows the measured surface topography. Aside from some contaminations, the sample surface is flat on a sub-nm scale. This can also be seen in the histogram in **Figure A6b.** For this sample area, a RMS surface roughness or $R_Q$ = 0.6 nm is found.

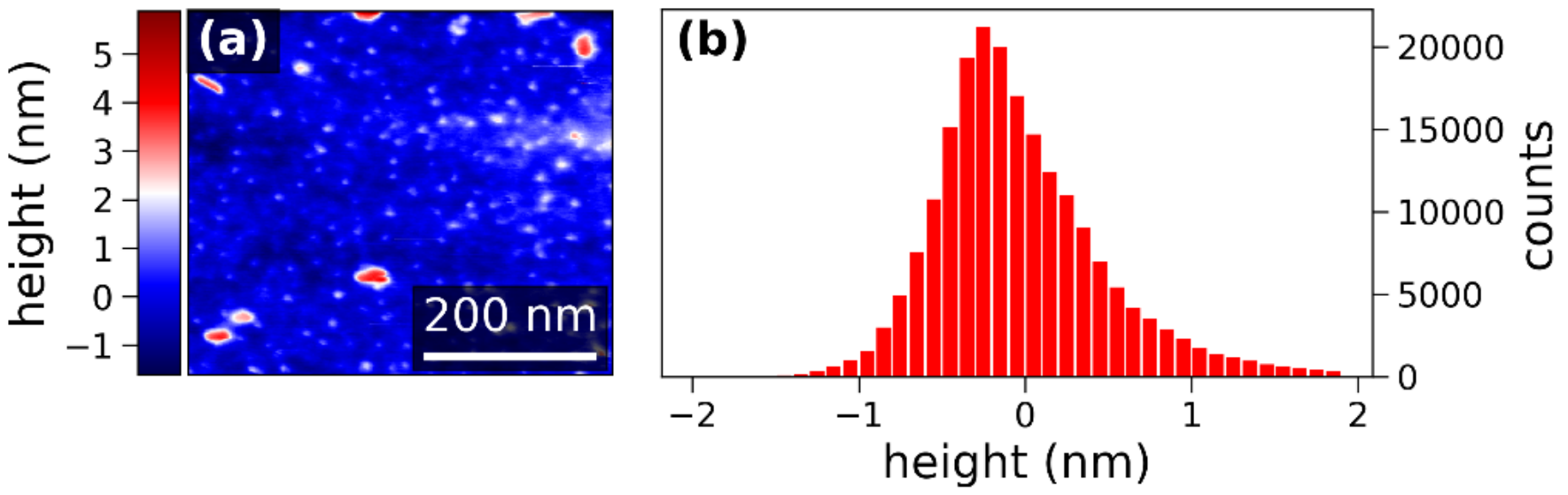

**Figure A5:** Surface roughness analysis: AFM topography image of the Ti/Pt/Co sample (a) and histogram plot of the height distribution (b).